\documentclass[twocolumn, floatfix, 
superscriptaddress, prb, 10pt]{revtex4-1}
\usepackage{extsizes}
\usepackage[paperwidth=210mm,paperheight=297mm,centering,hmargin=2cm,vmargin=2.5cm]{geometry}
\usepackage[utf8]{inputenc}

\usepackage{comment}
\usepackage{amssymb}

\usepackage{physics}
\usepackage{graphicx}%
\usepackage{dcolumn}%
\usepackage{bm}%
\usepackage{soul}
\usepackage{xcolor}
\usepackage{nicefrac}

\usepackage{times}
\usepackage{outlines}
\usepackage{chemformula}

\newcommand{\density}[3]{ \braket*{#1}{#2}_{#3} }
\newcommand{\densitydag}[3]{  \braket*{#2}{#1}_{#3} }

\newcommand{\densitysqt}[3]{ \braket*{#1}{#2^{(2)}}_{#3} }

\newcommand{\densitycube}[3]{ \braket*{#1}{#2^{(2)}}_{#3} }

\newcommand{\smearing}[1]{g\left(#1\right)}
\newcommand{\gaus}[2]{\CN_{#2}\pqty\big{#1}}
\newcommand{\ddelta}[1]{\delta(#1)}

\newcommand{\br}{\mathbf{r}}
\newcommand{\bri}{\mathbf{r}_i}
\newcommand{\brj}{\mathbf{r}_j}
\newcommand{\brij}{\mathbf{r}_{ij}}

\newcommand{\brhat}{\hat{\mathbf{r}}}

\newcommand{\bt}{\mathbf{t}}

\newcommand{\That}{\hat{t}}
\newcommand{\Rhat}{\hat{R}}

\newcommand{\dint}{\textrm{d}}

\newcommand{\dThat}{\dint\That}
\newcommand{\dRhat}{\dint\Rhat}

\newcommand{\CX}{\mathcal{X}}

\newcommand{\CA}{\mathcal{A}}

\newcommand{\CT}{\mathcal{T}}

\newcommand{\CN}{\mathcal{N}}
\newcommand{\sothree}{\ensuremath{\mathcal{SO}\left(3\right)}}

\def\CA{{\mathcal{A}}}

\setcitestyle{super}

\usepackage{hyperref}
\usepackage[capitalize]{cleveref}

\makeatletter
\AtBeginDocument{\let\LS@rot\@undefined}
\makeatother
\begin{document}

\title{Machine learning at the atomic-scale}

\author{F\'elix Musil}
\affiliation{Laboratory of Computational Science and Modeling, IMX, \'Ecole Polytechnique F\'ed\'erale de Lausanne, 1015 Lausanne, Switzerland}
\author{Michele Ceriotti}
\affiliation{Laboratory of Computational Science and Modeling, IMX, \'Ecole Polytechnique F\'ed\'erale de Lausanne, 1015 Lausanne, Switzerland}
\email{michele.ceriotti@epfl.ch}

\onecolumngrid
\begin{abstract}

Statistical learning algorithms are finding more and more applications in science and technology. Atomic-scale modeling is no exception, with machine learning becoming commonplace as a tool to predict energy, forces and properties of molecules and condensed-phase systems.
This short review summarizes recent progress in the field, focusing in particular on the problem of representing an atomic configuration in a mathematically robust and computationally efficient way. We also discuss some of the regression algorithms that have been used to construct surrogate models of atomic-scale properties. We then show examples of how the optimization of the machine-learning models can both incorporate and reveal insights onto the physical phenomena that underlie structure-property relations.

\end{abstract}
\twocolumngrid

\maketitle

\section{ Introduction}

The steady increase in computing power in the last decades, together with the improvements in accuracy and efficiency of electronic structure methods and empirical force fields (FFs), have given atomistic modeling a central role in the investigation of molecular and condensed-phase systems, and underpinned the rise of computational material design. 
Some recent achievements include the study of synaptic transmission mechanisms,~\cite{Song2018} water splitting with photo-electrical cells,~\cite{Pham2017} realistic metal deformations and plasticity~\cite{Zepeda-Ruiz2017} and nucleation with billions of atoms.~\cite{Shibuta2017}
Nevertheless the inherent scaling of \textit{ab initio} methods limits their applicability, preventing systems with more than a few thousand atoms from being studied, while the development of accurate and transferable reactive, multi-component empirical FFs remains a major challenge.
The last decade has seen the emergence of machine learning (ML) methods in the field of atomic-scale modeling to automate time consuming analyses~\cite{rohr+13arpc,rodr-laio14science,ceri19jcp} (unsupervised learning) or to reduce the cost of predicting quantities associated with atomic systems~\cite{Curtarolo2013, Ward2017, Mater2019, Faber2019} (supervised learning).
Unsupervised techniques aim at unravelling patterns in databases,  which in the context of atomistic modeling can correspond to identifying recurring motifs within structures,~\cite{Gasparotto2018,Helfrecht2019} as well as groups of `similar' structures in datasets of molecules and molecular solids~\cite{de+16pccp,musi+18cs} or molecular dynamics trajectories.~\cite{Blochliger2014,Kahle2019,Xie2019}
Given a set of atomic structures $\qty{\CA_n}$ associated with some properties $\qty{\bm{y}_n}$, e.g. energy or other observables computed by electronic structure theory, supervised ML methods can be used to learn a surrogate model $F:\CA\rightarrow\bm{y}$ to predict those properties. 
In this way, ML makes it possible to bypass solving Schr\"{o}dinger's equation, and to obtain inexpensive and accurate predictions of the formation energy of atomic structures,~\cite{rupp+12prl,Bartok2010} the chemical shieldings in molecular materials,~\cite{paru+18ncomm} the electron density of small molecules,~\cite{broc+17nc,gris+19acscs} the electron transfer coupling between dimers~\cite{Wang2019} etc.
One of the most promising applications for these algorithms is to provide frameworks to systematically build accurate interatomic potentials~\cite{Deringer2019, Behler2016, Qu2018} for a slightly higher running cost than traditional FFs.

In this review, we briefly summarize some of the approaches that have been used to model atomic scale properties with ML techniques. 
We begin by providing a detailed discussion of the problem of obtaining a representation of atomic configurations, i.e. how the Cartesian coordinates of the atoms can be transformed to obtain a mathematical description of the structure that is concise, and that incorporates the fundamental physical symmetry. In doing so, we will show how most of the existing representations can be seen as different views of a symmetrized atomic density. 
We then give a brief overview of the regression techniques that have been used in the context of atomic-scale modeling, focusing in particular on  Gaussian process regression, and discussing some of the aspects that are particularly relevant in the learning of atomic-scale properties. 
Finally we show how representations and regression models can be improved by incorporating more prior knowledge about the specific problem, using recent applications to highlight some of their key features. 

\section{Atomic-scale representations} \label{sec:rep}

The rise of ML during the last ten years has been mostly fueled by the emergence of models able to learn relevant features from raw data, e.g. images, texts, etc., alongside the parameters needed to perform tasks such as detecting objects or translating sentences.~\cite{Lecun2015}
In this context, deep-learning models that simply treat data as a stream - or an array - of bytes have outperformed models incorporating knowledge about the grammar of a language, or the content of a set of pictures.~\cite{Young2018}
Unlike many computer science applications, the properties of a physical system obey a number of symmetries and conservation laws, and efforts to encode these at the core of atom-scale models of matter have been shown to consistently improve the data efficiency of the regression scheme, make better use of the expensive electronic-structure calculations used for training.
One option is to incorporate symmetries at the level of the model. 
For example, extensions of the CNN architecture to extract invariant and/or covariant features from 3D shapes like an atomic structure~\cite{Cohen2018, Esteves2018, Thomas2018, Kondor2018, Weiler2018, Kondor2018-cgn} have been recently developed. 
The main approach followed in the atomic scale modeling community this far has however been to develop representations of the atomic structure that are equivariant with respect to these symmetries. 
Using these features as the input representation gives a ML model adapted to the desired symmetries.

Several authors have proposed to represent structures in terms of so-called fingerprints by concatenating features associated with an atomic structure, e.g. elemental properties, atomic connectivity, electronic structure attributes, stoichiometry, etc.~\cite{Ward2017,Seko2016,Mannodi-Kanakkithodi2016,Xue2016} to build models for complex properties such as melting temperature, dielectric constant and band gap energy.
While in principle any feature can be introduced into a ML model, electronic structure theory shows that any ground-state property of a structure $\mathcal{\CA}$ is a smooth function of the set of $N$ atomic coordinates $\left\{\br_i\right\}$ and chemical species $\left\{\alpha_i\right\}$.~\cite{Martin2004}
These considerations suggest that representations of the atomic structure based only on this core information provide a physically-motivated basis to regress any property $y(\CA)$ that could be computed by solving the Schr\"odinger equation for the structure. 

While a representation in terms of $\left\{\br_i,\alpha_i\right\}$ provides a complete description of a structure $\CA$, it does not incorporate the most basic physical symmetries that could follow a property, such as the invariance to the labelling of identical nuclei, or rigid translations and rotations of the reference frame. 
Many schemes have been proposed in recent years to translate the essential inputs of a quantum calculation code into a representation that incorporates these symmetries, and that can then be used in combination with most regression algorithms to learn physical properties in a data-efficient manner. 
Some start from internal coordinates of a molecule, such as the distances and angles between atoms,~\cite{behl-parr07prl,Braams2009,Xie2010,Qu2018,Shapeev2015,rupp+12prl,fabe+18jcp,hans+15jpcl} that are rotationally and translationally invariant while others begin with an atomic density~\cite{bart+13prb,Drautz2019,fabe+18jcp,Huo2017,Samanta2018,Seko2019,Hirn2017} which is invariant under the permutation of the atom indices.
As we illustrate below, many of these representations have been shown to be essentially equivalent, as they correspond to special cases of a general framework generating invariant and covariant representations from atomic densities.~\cite{will+18pccp, Grisafi2019chapter,Drautz2019} 
In the following text, we focus on local invariant representations but this framework is also a powerful tool to develop local covariant representations,~\cite{gris+18prl,will+18pccp,glie+17prb} as well as representations that capture non-local, global features of a given structure.~\cite{Grisafi2019-lode,de+16pccp}

We emphasize the generality and abstract nature of this construction by associating with each structure a vector $\ket{\CA}$. 
Different representations can be thought of as resulting from particular choices of the basis that is used to provide a concrete protocol to evaluate $\ket{\CA}$, much like the wavefunction can be expressed equally well in real space, in plane waves, or in one of the many localized basis sets that have been used in quantum chemistry. 
We choose a real-space basis as the starting point, and associate with $\ket{\CA}$ a set of element-resolved smooth atomic densities
\begin{equation}
    \density{\alpha \br}{\CA}{} = \sum_{i\in\CA,\alpha} \smearing{\br - \bri}. \label{eq:density}
\end{equation}
The sum extends over all atoms of type $\alpha$ within the structure, and $g$ is a smooth density function (a function peaked at zero with central symmetry that decreases to zero smoothly).
The use of a smooth density function instead of a Dirac distribution to represent the atomic coordinates ensures that the resulting representation is smooth with respect to atomic displacements.
Provided that the functions $g$ are sufficiently peaked, this representation determines fully the position of all the atoms, and is clearly independent on the order in which atoms are considered.

It is however not invariant with respect to rotations and translations. These additional symmetries can be incorporated through Haar integration~\cite{nachbin1976haar} of the atomic density, i.e. averaging over the corresponding group
\begin{equation}
    \ket{\CA}_{\hat{G}} = \int_{G}  \hat{G}\ket{\CA} \dint \hat{G},
\end{equation}
where $\hat{G}$ is an element of the group $G$. This averaging can be performed formally over the Dirac ket, but is more conveniently carried out by choosing a convenient basis in which to write explicitly the feature vector.
Furthermore, one should keep in mind that Haar integration -- just as any averaging procedure -- reduces the descriptive power of the representation. In other terms, structures that are distinct in terms of $\ket{\CA}$ might be indistinguishable when represented in terms of $\ket{\CA}_{\hat{G}}$.
For example a Haar integration of $\braket{\alpha\br}{\CA}$ over the translations $\That$ yields a constant scalar that counts the number of atoms of type $\alpha$ that are present in the structure.~\cite{will+19jcp} 
In order to avoid loss of resolving power, one can perform the average over tensor products of the atom density, i.e. evaluate the density at two different points and average over the simultaneous application of the symmetry operation to both points. 
To be concrete, let us derive explicitly this representation for a Gaussian smearing function $\gaus{\br}{\sigma^2} = \exp(-\br^2/2\sigma^2)$. 
To retain structural information, we compute a translationally-symmetrized representation based on a two-point evaluation of the atom density:
\begin{align}
    \densitysqt{\alpha \br\alpha'\br'}{\CA}{\That} = & \sum_{\substack{i\in\CA,\alpha\\j\in\CA,\alpha'}}   \int_{\mathbb{R}^3}\dThat \begin{aligned}[t]    \Big[\gaus{\That\br'-\brj}{\sigma^2} \\ \gaus{\That\br -\bri}{\sigma^2} \Big] \end{aligned} \nonumber \\
    =& \sum_{\substack{i\in\CA,\alpha\\j\in\CA,\alpha'}} \int_{\mathbb{R}^3}\dint \bt \begin{aligned}[t]  
    \Big[\gaus{\br'+\bt-\brj}{\sigma^2} \\
        \gaus{\br + \bt -\bri}{\sigma^2} \Big] \end{aligned}
     \nonumber \\
    = &\sum_{\substack{i\in\CA,\alpha\\j\in\CA,\alpha'}} \gaus{\br - \br' -\bri + \brj}{2\sigma^2} \nonumber\\
\Rightarrow \densitysqt{\alpha\alpha'\br}{\CA}{\That} = & \sum_{\substack{i\in\CA,\alpha\\j\in\CA,\alpha'}} \gaus{\br -\brij}{2\sigma^2},
\label{eq:sym-trans}
\end{align}
where $\br_{ij}=\brj-\bri$ and $\br - \br'$ has been replaced with $\br$.
Note that translational averaging reduced by three the number of independent variables, and that the symmetrized density takes the structure of a many-body expansion of the potential energy truncated up to the pair contributions.
In other words a symmetrized pair-density representation of an atomic structure can be decomposed into a sum of representations centered on each of the atoms.
Moreover while one could consider all the pairs in the representation, the nearsightedness principle of electronic matter,~\cite{prod-kohn05pnas} which underlies most linear-scaling electronic structure methods,~\cite{gall-parr92prl,goed99rmp, Papadopoulos2011, Bowler2012} and the clear computational advantage deriving from restricting the range of atomic pairs that need to be included in the sum, motivates the limitation of the atomic neighborhood to a sphere of radius $r_c$ centered on each atom through a cutoff function $f_c(r)$ that is zero for $r>r_c$.
To simplify notation, we then introduce an atom-centered symmetrized density representation
\begin{equation}
    \density{\alpha\br}{\CX_i}{} =  \sum_{j\in\CX_i,\alpha}  \; \gaus{\br -\brij}{2\sigma^2} f_c(r_{ij}), \label{eq:trans-density}
\end{equation}
where $\CX_i$ is an atomic environment centered on atom $i$ that includes all the neighbors within a sphere of radius $r_c$.
The cutoff function should smoothly decay to zero to avoid introducing a discontinuity with respect to atoms entering/leaving the atomic neighborhood in the representation.
Using this notation, one can write 
\begin{equation}
\density{\alpha\alpha'\br}{\CA^{(2)}}{\That}= \sum_{i\in\CA,\alpha'} \density{\alpha\br}{\CX_i}{}.
\end{equation}

The environment-centered features $\ket{\CX_i}$ are not rotationally invariant, and so one can proceed to the symmetrization over the rotation group. Using the z-y-z Euler parametrization, one can compute
\begin{equation}
\begin{split}
    \density{\alpha\br}{\CX_i^{(1)}}{\Rhat} = & \sum_{j\in\CX_i,\alpha}  f_c(r_{ij}) \int_{\text{SO(3)}} \gaus{\Rhat\br - \brij}{2\sigma^2} \dRhat \\
    =& \begin{aligned}[t] 2\pi \sum_{j\in\CX_i,\alpha}  f_c(r_{ij}) \int_{0}^{\pi}\sin{\theta}\dint\theta  \int_{0}^{2\pi} \dint\phi \\ \exp[-\frac{r^2+r_{ij}^2 - 2rr_{ij}\cos{\theta}}{4\sigma^2} ]   \end{aligned} \\
    =& \begin{aligned}[t]8\pi^2 \sum_{j\in\CX_i,\alpha}  f_c(r_{ij})\sinh[rr_{ij}/2\sigma^2] \\ (rr_{ij}/2\sigma^2)^{-1} \exp[-(r^2+r_{ij}^2)/4\sigma^2]\end{aligned} \\
    \density{\alpha r}{\CX_i^{(1)}}{ \Rhat} \approx& \sum_{j\in\CX_i,\alpha}  f_c(r_{ij})r_{ij}^{-1} \gaus{r-r_{ij}}{2\sigma^2} , %
    \label{eq:2b-sym}
\end{split}
\end{equation}
where we note that the integration makes the orientation of ${\br}$ irrelevant, and we write the feature vector as a function of $r=\norm{\br}$. 
Some constant factors and the $ \gaus{r+r_{ij}}{2\sigma^2}$ term have been omitted because they do not contribute to the representation since $r,r_{ij}>0$ with $\sigma$ relatively small.  
Note also that we have introduced in the definition of $\density{\alpha r}{\CX_i^{(1)}}{ \Rhat}$ an additional factor of $r$, so that
\begin{equation}
\begin{split}
    \int_{\mathbb{R}^3}\densitydag{\alpha \br}{\CX_i^{(1)}}{\Rhat} &\density{\alpha\br}{\CX_j^{(1)}}{\Rhat}\dint \br =  \\
    &\int_0^{\infty} \densitydag{\alpha r}{\CX_i^{(1)}}{ \Rhat} \density{\alpha r}{\CX_j^{(1)}}{ \Rhat}\dint r .
\end{split}
\end{equation}
This symmetrized density $ \density{\alpha r}{\CX_i^{(1)}}{ \Rhat}$ is essentially a 2-body correlation function resulting from a Gaussian kernel density estimation (KDE).
The body order naturally characterizes the amount of information included in an invariant density representation.

\begin{figure*}[bhtp]
\centering
\includegraphics[width=1.0\linewidth]{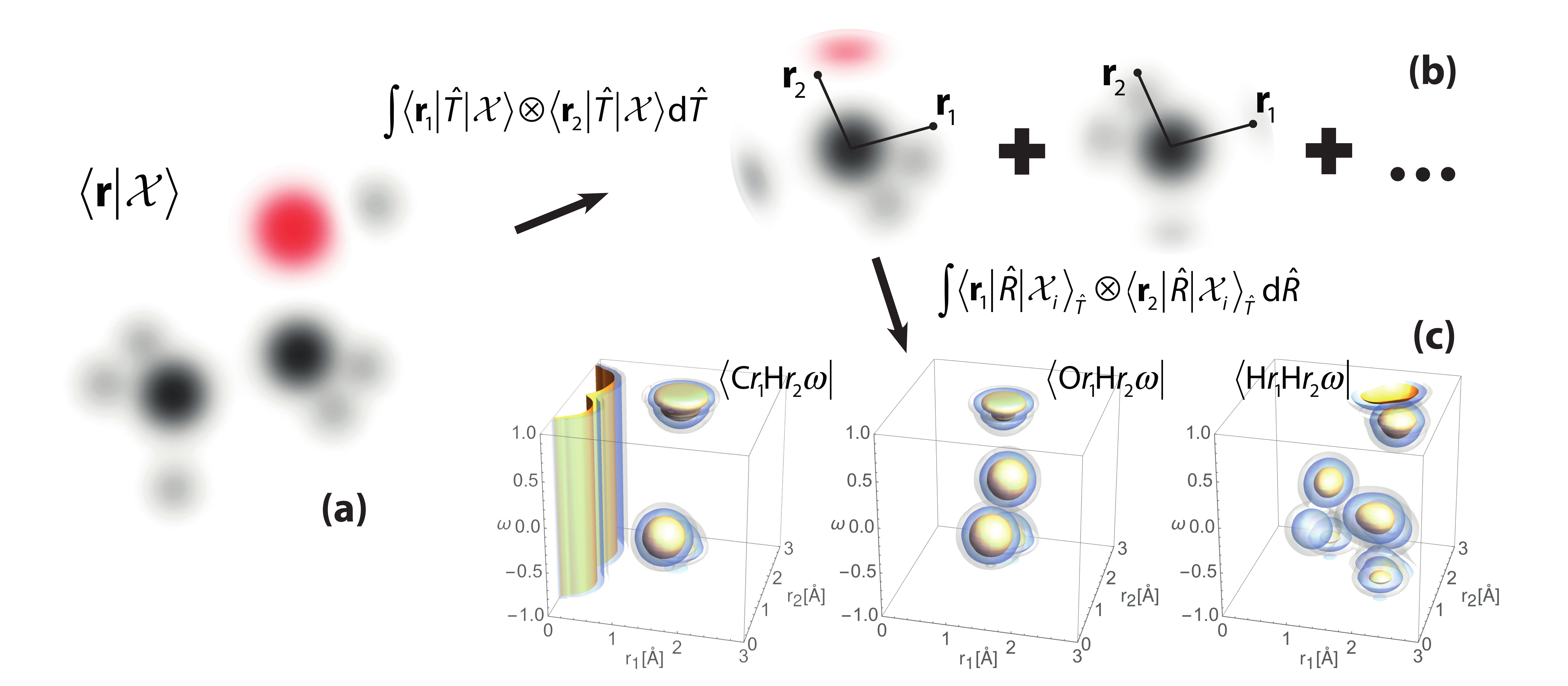}
\caption{A graphical summary of the steps leading from a decorated atomic density to the $3$-body invariant representation of ethanol in real space $\ket*{\CX_i^{(2)}}_{\Rhat}$. (a) The geometry of a small molecule is mapped into a smooth atom density using a Gaussian smearing function. The chemical composition represented by the elemental ket $\ket{\alpha}$ is color coded: carbons are black, oxygen is red and hydrogens are grey. (b) The symmetrization over the translational group of a two-point density results in the decomposition of the representation into a sum of atom centered contributions where a finite cutoff has been applied (see \cref{eq:trans-density}). (c) The symmetrization over the rotational group with $\nu=2$ delivers the $3$-body invariant representation. Some isocontours of $\densitycube{\alpha r_1 \beta r_2 \omega}{\CX}{\Rhat}/r_1r_2$ associated with the central carbon atom illustrate some of the real space features extracted with the atom density framework. Adapted from Ref.~\citenum{will+19jcp}. }
\label{fig:density}
\end{figure*}
It is clear that this procedure is very general, and can be applied to any tensor power of the density, both when integrating over translations and when integrating over rotations. 
Increasing the order $\mu$ of the product in the integration over $\That$ leads to $\mu-1$ nested sums over the atomic neighborhood which might not be computationally favorable.
Increasing the order $\nu$ of the tensor product in the integral over the continuous rotation group similarly increases the body order of the structural correlations described by $\ket*{\CX_i^{(\nu)}}_{\Rhat}$. 
If one did so while writing explicitly the environmental ket $\ket{\CX_i}$ as a sum over neighbors, this procedure would increase the order of the sum over neighboring atoms. 
One can however also proceed by expanding $\ket{\CX_i}$ in an appropriate basis, e.g. a combination of radial functions $R_n(r)$ and spherical harmonics
\begin{equation}
\density{\alpha nlm}{\CX_i}{} =
\int\mathrm{d}\br\  R_n(r) Y_l^m(\brhat) \density{\br}{\CX_i}{},
\label{eq:sph-exp}
\end{equation}
in which case higher-order invariants can be written as sums over the expansion coefficients,
\begin{equation}
    \density{\alpha n \alpha' n'l}{\CX_i^{(2)}}{\Rhat} = \frac{1}{\sqrt{2l + 1}}\begin{aligned}[t] \sum_m (-1)^m \density{\alpha' n'lm}{\CX_i}{} \\ \density{\alpha nl-m}{\CX_i}{}  . \end{aligned}
\end{equation}
The flexibility of this framework allows links to be drawn between several representations that  might otherwise look quite dissimilar.
The type of smearing function used to construct the atomic density, the basis onto which the density is represented (real space grid, orthonormal basis set, etc.), can impact the effectiveness and the computational efficiency of the resulting implementation but do not change the fundamental nature of the invariant representation.
For example, the choice of Gaussian smearing and a basis of radial functions corresponds to the smooth overlap of atomic positions (SOAP) framework,~\cite{bart+13prb,de+16pccp} with the power spectrum and the bispectrum  corresponding to rotational averages with $\nu=2$ and $\nu=3$ respectively.
The computation of these coefficients involves the evaluation of several costly special functions.~\cite{Grisafi2019chapter}
Even if the cost of evaluating SOAP features can be reduced greatly by the introduction of approximations and numerical workarounds,~\cite{Caro2019} the use of both a smooth atom density and a smooth basis set might seem redundant and costly.
This led \citet{Drautz2019} to use Dirac distributions in the representation of the density, and obtain smoothness by truncating the basis set on which this density is expanded.
The resulting invariant representations correspond precisely to the $g\rightarrow\delta$ limit of the SOAP power spectrum, bispectrum and higher-$\nu$ invariants, but can be expressed in terms of simpler mathematical functions.

The expansion on a complete basis set of the atomic density ensures the general applicability of a representation but it also increases its computational cost by probing regions of the configurational space that are not relevant for a given system.
The symmetry functions~\cite{behl-parr07prl} framework make it possible to use the knowledge of the system at hand to carefully tailor a representation of the atomic environment.
The resulting representation can be interpreted as a projection on these symmetry functions fixed in particular regions of the configurational space with the $\delta$-limit of the $(\nu+1)$-body invariant ket, 
\begin{equation}
    \density{\alpha G_2}{\CX_i}{} =  \int \dint r \,G_2(r) \density{\alpha r}{\CX_i^{(1)}}{\Rhat,g\rightarrow\delta},
\end{equation}
where $\density{\alpha r}{\CX_i^{(1)}}{\Rhat,g\rightarrow\delta} =  \sum_{j\in\CX_i,\alpha} \; \ddelta{r -r_{ij}} f_c(r_{ij})$ and $\delta$ is the Dirac distribution.
Other recently-introduced feature vectors for atomistic learning, such as the FCHL~\cite{fabe+18jcp} and the MBTR~\cite{Huo2017} representations, use an adaptive basis to smooth the $\delta$-limit of the $(\nu+1)$-body invariant ket $\ket*{\CX_i^{(\nu)}}_{\Rhat}$, effectively constructing a kernel density estimate of the invariant correlation function.
These two representations differ by the choice of kernel functions and how they encode the chemical information, with the FHCL features using a kernel to encode the similarity between different elements, similarly to what was done in Ref.~\citenum{de+16pccp}.

The density based representation framework makes it possible to rigorously formulate a hierarchy of invariant representations and shows that several commonly used descriptions of atomic structures and environments actually contain a similar amount of information.
This formal connection is also reflected in several recent extensive empirical benchmarks,~\cite{Nguyen2018, Nyshadham2019, Zuo2019} that show that many of these representations actually perform similarly in terms of model accuracy, while the main difference between them is their computational cost.

It is also worth mentioning that density-based invariants can be generalized to yield feature vectors of the form $\ket{\CX^{(\nu)}\lambda\mu}_{\Rhat}$ that transform covariantly under rotations of the reference frame as the spherical harmonics $Y_\lambda^\mu$,~\cite{gris+18prl,Grisafi2019chapter} providing a symmetry-adapted basis to learn  properties such as atomic forces, elastic moduli or dielectric response tensors, which also rotate rigidly under $\sothree$ group operations. 
Using a generic atom-centered symmetry-adapted representation has proven to be more effective~\cite{Raimbault2019} than frameworks that assume a rigid molecular frame to achieve covariance of the predicted properties,~\cite{Bereau2015,liang2017} and has made it possible to learn an atom-centered decomposition of a scalar field like the electron density.~\cite{gris+19acscs,Fabrizio_2019}
Although learning schemes based on covariant features or kernels~\cite{glie+17prb,Botu2015,li+15prl,Mailoa2019} could in principle be used to machine-learn directly the inter-atomic forces rather than the underlying atomic potential, enforcing energy conservation has proven difficult. For this reason, most of the existing machine-learning interatomic potentials are built to predict the potential, although they can incorporate forces as an indirect learning target.~\cite{behl-parr07prl,Bartok2010,Ceriotti2018,Chmiela2017,Schutt2018}

\section{Machine learning quantum mechanics}

ML algorithms for regression~\cite{Bishop2006} aim to construct a model $y=F(\CA)$ that can predict accurately the properties of a structure. 
The internal parameters of the model are determined by optimizing the accuracy of prediction over a set of training structures, $\left\{\CA_i,y_i\right\}$, and their accuracy with respect to that reference can be improved systematically by increasing the size of the training set.~\cite{Amari1993}
One of the early applications of ML to the prediction of atomic-scale properties aimed at obtaining an accurate model of the potential energy surface (PES), which is crucial to assess the stability of a given configuration, and whose sampling underlies the evaluation of the thermodynamic properties of a system.~\cite{TuckermanMark1972}
Contrary to traditional FFs, which assume physics-inspired functional forms for the interactions, and often use experimental observable as fitting targets, ML interatomic potentials (MLIPs) don't assume a fixed functional form, and usually rely on electronic-structure calculations as a reference. 
In many cases, this more general, data-driven approach has been shown to result in more transferable and accurate models.~\cite{Bartok2010, Behler2007, Braams2009, rupp+12prl}
Besides the PES, ML models have also been successful at predicting other zero Kelvin properties such as chemical shieldings, band gaps, electron affinities, electron transfer integrals and static isotropic polarizabilities.~\cite{paru+18ncomm,mont+13njp,de+16pccp,musi+18cs,Schutt2018,fabe+17jctc,bart+17sa,Homer2019} 
While considerable success has also been shown in using ML to predict complex properties that cannot be seen as arising from an individual atomic configuration (e.g. the free-energy of a state, the toxicity or pharmaceutical activity of a molecule, etc.), here we will focus entirely on the well-defined task of building a surrogate quantum model, which can sidestep the solution of the Schr\"odinger equation and predict the properties of a specific atomic configuration.
In this section we summarize the regression methods that have been applied to perform such prediction. 
While the main focus will be on the construction of interatomic potentials, we will keep the discussion as general as possible, and mention how the different approaches should be modified to deal with other classes of properties. 

A scalar property $y(\qty{\br_i,{\alpha_i}})$ of a system $\CA$ of $N$ atoms of species $\alpha_i$, located at positions $\br_i$,  can be expressed formally as a function of an abstract vector of features $\ket{\CA}$ that represents the structure,
\begin{equation}
    y(\CA) = F(\ket{\CA}).
\end{equation}
The problem of modeling $y(\CA)$ can therefore be decomposed into the problem of providing a concrete formulation of the feature vector (that we have discussed in detail in the previous Section) and that of determining the functional form of the approximating model $F$. 
Irrespective of the regression technique used, most of the transferable property models that have been introduced in recent years decompose a property associated to a set of atoms $\CA$ into atom-centered contributions, i.e.
\begin{equation}
    y(\CA) = \sum_{i \in \CA} f(\ket{\CX_i}), \label{eq:additive-model}
\end{equation}
where $f$ is a trained ML model and $\CX_i$ indicates the atomic environment centered on atom $i$ of structure $\CA$.
This choice can be motivated as a consequence of imposing the invariance of the property on the absolute position of the system (see \cref{eq:sym-trans}), and -- together with the limitation of the range of each environment to a region centered on the $i$-th atom -- yields models of great transferability, since it allows breaking down the properties of large, complex configurations into a sum of contributions that only depend on the position of a few dozen atoms.
In the cases in which this ansatz is not  justified (e.g. for properties such as ligand binding affinity, or in the presence of significant long-range interactions) other strategies for combining local environments predictions like the REMatch kernel should be considered.~\cite{bart+17sa}

Linear models based on permutation invariant polynomials (PIPs) have been very effective at reproducing accurately chemical reactions between small molecules~\cite{Braams2009,Xie2010,Qu2018} and to build efficient MLIPs with the many-body tensor (MBT) framework~\cite{Shapeev2015} that extends them to more complex systems.~\cite{Jafary-Zadeh2019,Novoselov2019}
Similarly linear models based on the $n$-body correlation function~\cite{Thompson2015,Wood2018,Drautz2019,Seko2019,Deng2019} have shown great promise.
Fully non-linear models based on artificial neural networks (ANN) have however been the most popular this far. ANNs have been constructed based on the the expansion of the radial (and angular) distribution function on a basis such as the Behler-Parrinello symmetry functions~\cite{Behler2007,Behler2011,Artrith2016,Smith2017,Yao2018,Lee2019,Herr2019,Zhang2018}, Zernike polynomials~\cite{Khorshidi2016}, Chebychev polynomials\cite{Artrith2017}, Gaussians~\cite{Schutt2017,Schutt2018,Lubbers2018,Unke2019}, and proved very successful at investigating the properties of complex systems.~\cite{Natarajan2016,Gastegger2017,Kapil2016,Cheng2019, Huang2018,Eckhoff2019,Homer2019}
Another class of models that have been both very popular and successful is based on Gaussian process regression (GPR),~\cite{Rasmussen2006} that is formally equivalent to kernel ridge regression (KRR) and can be seen as a middle-ground solution that introduces non-linearity in the form of a kernel function $k(\CX,\CX')$ built on pairs of feature vectors, but effectively translates into a linear regression problem that uses (some of) the training set structures as the basis on which the structure-property relation is constructed.
GPR has been used to predict the stability of molecules and solids~\cite{rupp+12prl,fabe+17jctc,Ramakrishnan2015,Hansen2013,fabe+16prl,bart+17sa,de+16pccp,de+16jci,musi+18cs,Bartok2010,glie+18prb} and build MLIPs for elemental solids,~\cite{szla+14prb,deri-csan17prb,Deringer2018-silicon,Bartok2018} nano clusters,~\cite{Zeni2018} isolated molecules~\cite{Chmiela2017} and molecular liquids~\cite{Veit2018} as well as for the direct prediction of other quantum mechanical properties.~\cite{Raimbault2019,musi+18cs,paru+18ncomm,physrevb.89.235411,mont+13njp,Ramakrishnan2015-jcp,Christensen2019}

In the most straightforward form, a GPR model built on a kernel function $k$ can be written based on a set of $N$ training structures $\qty{\CT_n}$, and the associated properties $y_n$.
Assuming a Gaussian likelihood, and an additive, atom-centered property model, the prediction for a structure $\CA$ becomes
\begin{equation}
    y(\CA) =  \sum_{n=1}^N x_n K_{\CA\CT_n}, 
\end{equation}
where $K_{\CA\CT_n} = \sum_{i \in \CA} \sum_{j\in \CT_n} k(\CX_i,\CT_{n,j})$ and the kernel function $k(\cdot,\cdot)$ quantifies the similarity between the local environments of $\CT_n$ and the centered structure $\CX_i$.
The key ingredient of this model is the kernel function that - subject to a few conditions such as positive definiteness - defines an inner product between the inputs $k(\CX_i,\CX_j)$.  
The repesenter theorem~\cite{Scholkopf2001} guarantees that the kernel can be associated with an inner product between vectors in a Hilbert space, i.e.  $k(\CX_i,\CX_j)=\braket{\CX_i}{\CX_j}$. 
The use of the Dirac notation underlines the independence of the basis, i.e. representation or features, used to effectively quantify the similarity between atomic configurations. 
In some cases - for instance in the case of the SOAP representation discussed in the previous Section - it may be possible to write explicitly the feature vectors associated with a given kernel.

GPR is often preferred over the more sophisticated non-linear models because of its ease of use: it has a single interpretable hyperparameter $\sigma$, and the solution for the weights $x_n$ has the closed form
\begin{equation}
    \bm{x} = \bm{K}^{-1} \bm{y},
    \label{eq:gpr-training}
\end{equation}
where $K_{nm} =  K_{\CT_n\CT_m} + \sigma^2 \delta_{nm}$ is the kernel matrix between the $N$ training inputs and $\bm{y}_n$ is the property associated with structure $\CT_n$;
$\sigma$ corresponds to an expected Gaussian noise in the references $\bm{y}$ so it can account for small discrepancies in the convergence of the electronic structure method that are often found across a training set.
In the language of kernel ridge regression, \cref{eq:gpr-training} can be obtained by minimizing the loss
\begin{equation}
\mathcal{L}(\bm{x}) = \sum_n \left|y(\CT_n)-y_n\right|^2 + \sigma^2 \sum_n x_n^2.
\label{eq:loss}
\end{equation}
It should be mentioned that GPR provides a simple approach to compute derivatives of the target properties with respect to atomic positions, e.g. the force consistent with the model, in which case $y$ represents the PES of a configuration.
Such derivatives are easily expressed in terms of derivatives of the kernel, i.e.
\begin{equation}
    \frac{\partial y(\CX_i)}{\partial \br_i} =  \sum_{n=1}^N\sum_{j\in \CT_n} x_n  \frac{\partial}{\partial\br_i} k(\CX_i,\CT_{n,j}). %
\end{equation}
Derivatives can also be incorporated in the learning procedure,~\cite{Rasmussen2006,Christensen2019,Bartok2015,Ceriotti2018,Chmiela2017,glie+18prb} by including the discrepancy between reference and predicted values in the loss \cref{eq:loss}. 
Building a symmetry-adapted GPR model for properties that have a tensorial nature requires the construction of covariant kernels,~\cite{glie+17prb,gris+18prl} that describe the correlations between the spherically-covariant components of the target property, 
\begin{equation}
    y^\mu_\lambda(\CA) = \sum_{n=1}^N\sum_{m=-\lambda}^\lambda x_{nm} \left[K_{\CA\CT_n}\right]^\lambda_{m\mu}. 
\end{equation}
For instance, a kernel which fulfills these symmetry requirements can be constructed based on $\lambda$-SOAP features,~\cite{gris+18prl}
\begin{equation}
k^\lambda_{\mu\mu'}(\CX,\CX') = \sum_{nn'll'} \braket{\CX^{(2)}\lambda\mu}{nn'll'}\braket{nn'll'}{{\CX'}^{(2)}\lambda\mu}.
\end{equation}
Finally, the probabilistic nature of GPR also allows one to estimate the uncertainty associated with the prediction
\begin{equation}
    \sigma_y(\CA) = \sigma^2 + K_{\CA\CA} -  \bm{K}_{N \CA }^T\bm{K}^{-1}\bm{K}_{N \CA}.
\end{equation}

The drawback for such simplicity is the computational cost associated with the training phase - which scales cubically with the training set size -  and the need to use the full training set as a basis to perform predictions.
To address this issue, many approximations of the exact kernel matrix have been proposed,~\cite{Quinonero-Candela2005, Liu2018} among which the projected process (PP) approximation ~\cite{Seeger2003, Quinonero-Candela2005} has been shown to be quite practical to include force references~\cite{Bartok2015, Ceriotti2018} and effective from the point of view of the cost and accuracy of predictions.~\cite{Bartok2018, Musil2019}
The PP method introduces $M$ pseudo inputs to approximate the GP prior which practically reduces the cost of training to the inversion of a $M\times M$ matrix, and ensures that predictions only require computing kernels between the new configurations and the $M$ pseudo inputs:
\begin{equation}
    \begin{split}
        y^{\mathrm{PP}}(\CA) &=  \bm{K}_{M\CA}^T \tilde{\bm{K}}^{-1} \bm{K}_{MN} \bm{y}, \\
        \sigma^{\mathrm{PP}}_y(\CA) &= \begin{aligned}[t]\sigma^2 + K_{\CA\CA} - \bm{K}_{M\CA}^T \bm{K}_{MM}^{-1} \bm{K}_{M\CA} \\ + \bm{K}_{M\CA}^T \tilde{\bm{K}}^{-1} \bm{K}_{M\CA},\end{aligned}
    \end{split}
\end{equation}
where $\tilde{\bm{K}} = \bm{K}_{MM} + \sigma^{-2}\bm{K}_{NM}^{T}\bm{K}_{NM}$, $\bm{K}_{MM}$ indicates the kernel matrix between pseudo inputs, and $\bm{K}_{NM}$ the matrix between training points and pseudo inputs.
For simplicity, the pseudo inputs (or active points) can be chosen directly from the training set and they represent a new basis in which the regression is performed.
To maximize the cost reduction and the accuracy of the model, one needs to sample the active set carefully.
Selecting randomly the active inputs is far from optimal so several approaches have been proposed~\cite{Smola2001,Seeger2003,SathiyaKeerthi2005,Schreiter2016} among which Farthest Point Sampling (FPS),~\cite{ceri+13jctc} a greedy method that maximises diversity, or a CUR decomposition~\cite{Bartok2015,imba+18jcp} of the feature matrix associated with the training set, which minimizes the effect of the PP on the kernel matrix, have allowed significant reductions of the computational cost with minimal degradation of the accuracy.~\cite{Bartok2018, Musil2019}

\begin{figure}[bhtp]
 \centering
 \includegraphics[width=0.45\textwidth]{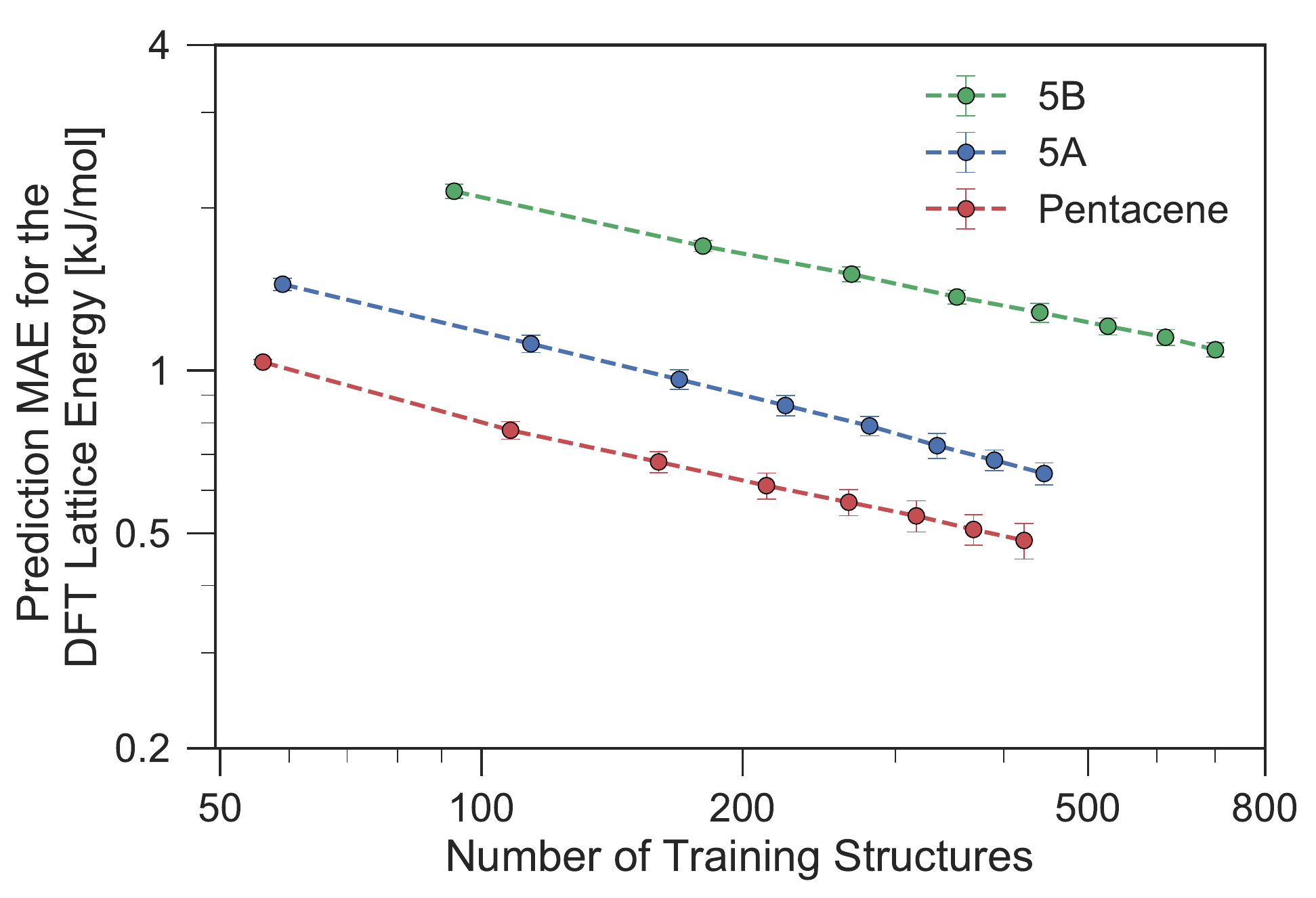}
    \caption{Learning curves for the lattice energy predictions of pentacene, 5A and 5B datasets, plotted on on a logarithmic scale. 
    For each training sample size, models are built several times on random subsets of the full training set and predictions are made on a fixed-size random subset of the training set.
    The test MAE and error bars are, respectively, average and standard deviation over the random subset predictions.
    All hyper-parameters of our ML model are fixed except for the regularization parameter $\sigma$ in the GPR model which is optimized on the fly at each training. Adapted from Ref.~\citenum{musi+18cs}.}
 \label{fig:learning-curve}
 \end{figure}

ML algorithms include recipes to train their parameters, e.g. \cref{eq:gpr-training}, but they do not specify how to determine hyperparameters such as the regularization $\sigma$ for GPR, the number of layers in an ANN and the cutoff radius $r_c$ in the power spectrum representation, which can influence heavily the quality of the model.
In the Bayesian context these hyperparameters can be interpreted as priors that should be inferred from our knowledge of the physical system,~\cite{Bartok2015} or thought of as parameters that need to be optimized.
In principle the best parameters should allow for the lowest possible prediction error on all possible inputs.
Given that one can only work on a finite-sized set of references, the problem becomes to find the parameters that best reproduce the available references and at the same time generalize well to unknown inputs.
The performance of a model is measured by comparing the predicted values and the reference values with metrics such as the mean absolute error (MAE) or the root mean square error (RMSE).
An effective technique to avoid overfitting these parameters, i.e. specialize the model for the training set which leads to poor generalization performances, is the so called $k$-fold cross-validation where the performances are evaluated on several subsets of the training set (see \citet{Hansen2013} for more details).
Cross validated scores are more likely to match the generalization error which is a good basis to rank models and determine the optimal set of hyperparameters.~\cite{VonLilienfeld2018}
Learning curves are another standard diagnostic tool to characterize the performance of ML models.
From statistical theory, the error of a given model decreases as a power-law with the size of the training set.~\cite{Amari1993}
\Cref{fig:learning-curve} shows, on a logarithmic scale, three learning curves for models trained on datasets of molecular crystal polymorphs to reproduce their lattice energies.
The GPR model performances vary with the considered training set because the learning rates (slopes of the curves) and off-sets are different.
These curves are very useful because they help differentiating between models that have a small offset and learning rates with models that have a larger off-set but also steeper slopes (see \cref{fig:elpasolites-lc} for an example).
Indeed, building a `good' model with as few references as possible might be favored over a model that has a better learning power but poorer performances with few samples.

 \begin{figure*}[bhtp]
 \centering
 \includegraphics[width=1.0\textwidth]{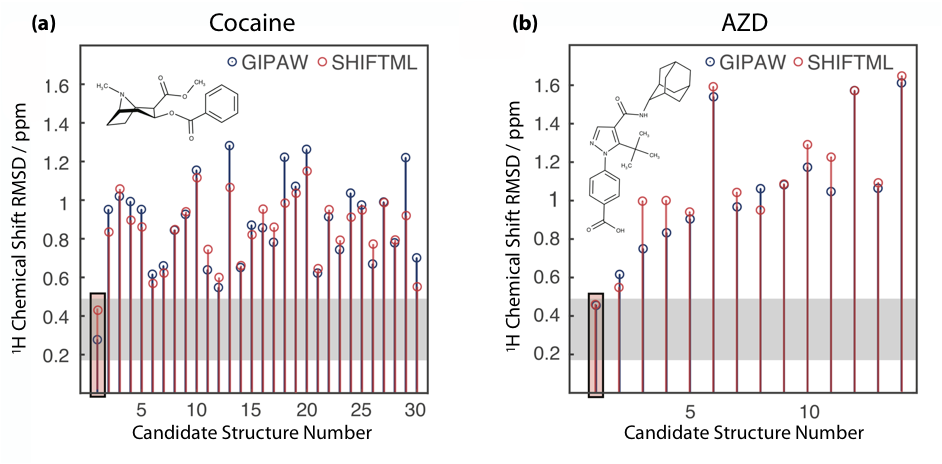}
    \caption{Structure determination for cocaine (a) and AZD8329 (b) obtained by comparing calculated and experimental $^1$H chemical shifts for the most stable structures obtained with CSP. 
    The total RMSE between experimentally measured shifts (NMR spectroscopy) and shifts calculated with GIPAW~\cite{Pickard2001,Yates2007} (blue) and ShiftML~\cite{paru+18ncomm} (red) is shown for every hypothetical structure. The shaded area represents an estimation of the confidence intervals for the total RMSE computed with GIPAW. The candidates that have RMSEs within this range are the most likely observed crystal structures using a chemical shift-based solid-state NMR crystallography protocol.~\cite{Salager2010}  Adapted from Ref.~\citenum{paru+18ncomm}.}
 \label{fig:csd-sd}
 \end{figure*}

Even though learning curves and cross-validation procedures can benchmark quantitatively the ability of a model to perform well in production, demonstrating the performance of a model on practical test cases is typically more compelling.
For example, \cref{fig:csd-sd} shows how the ShiftML model for the $^{1}$H chemical shifts~\cite{shiftml} is able to identify the crystal structure observed experimentally with NMR spectroscopy of two molecular materials as well as GIPAW DFT, the reference method used to train it.
In the better-established case of the construction of MLIPs, several recent works have started to compare systematically the ability of different schemes to reproduce \textit{ab initio} energies and forces,~\cite{Li2018} the short range interaction in the MB-pol water model,~\cite{Nguyen2018} the vibrational spectra of H$_2$CO,~\cite{Kamath2018} the radial and angular distribution functions of copper and silica and the equation of state of three binary alloys.~\cite{Zuo2019}
Overall these studies show that all of the models considered were able to reproduce observables within the expected accuracy of the underlying electronic structure reference.
In light of the substantially equivalent asymptotic accuracy of different approaches, the preference for one MLIP over another depends more on practical considerations such as training data efficiency, computational cost, simplicity of use, etc.

The ability of a ML model to reproduce the results of reference calculations on a validation/test set makes it possible to assess its overall quality, but it does not guarantee that the predictions are equally accurate.
A reliable uncertainty estimate that provides an assessment of the model accuracy \emph{for a specific prediction} is key to allow for a wider community of researchers to rely on ML models. %
A punctual quantification of ML uncertainty is also useful as a criterion for the iterative improvement of a model's training set with active learning~\cite{Smith2018, Gubaev2018, Vandermause2019,Jinnouchi2019}, as one would like to incorporate additional reference data in the regions that correspond to the least accurate predictions.
Several techniques such as GPR, Bayesian neural networks (BNN)~\cite{Mackay1995, Blundell2015} and ensemble models~\cite{Dietterich2000} have been developed to provide an estimate of the uncertainty associated with predictions.
Model ensembles, which estimate uncertainty by performing multiple predictions for each input, have been quite popular~\cite{pete+17pccp,Behler2017,Smith2018,Musil2019} because of their simplicity and flexibility.
The resampling approach in particular~\cite{Efron1979,Politis1999,Politis1994,Tibshirani1996} is based on the training of a family of models, $N_r$ different subsets of the training data. 
It produces a non-parametric estimate of the predictions distribution $P(y|\CA)$ whose moments are given by
\begin{equation}
    \begin{split}
        \bar{y}(\CA) &= \frac{1}{N_r} \sum_i y^{(i)}(\CA) \\
        \bar{\sigma}^2(\CA) &= \frac{1}{N_r -1} \sum_i [y^{(i)}(\CA)-\bar{y}(\CA)]^2,
    \end{split}
    \label{eq:committee-model}
\end{equation}
where $y^{(i)}(\CA)$ is the prediction of the $i^{th}$ resampled model.
While the training cost is increased $N_r$ times, uncertainty predictions come with the estimate of $y(\CA)$ at essentially no extra cost for GPR -- which is typically dominated by the evaluation of the kernel.  
In the case of ANN, an ensemble of models provides a practical way of estimating the uncertainty, although in this case the overhead can be significant, and linear in $N_r$.
To avoid such overhead, as well as the increased training cost, dedicated schemes that avoid training multiple models have been developed specifically for ANN.~\cite{Segu2019,Janet2019,Ryu2019}

\section{Optimizing the Representations} \label{sec:opt-rep}

As discussed previously, representations of an atomic structure for atomic scale simulations should provide a concise but complete description of its structure and composition. 
Ensuring that these features follow the basic symmetries of the target property is an essential condition, but does not guarantee optimal performance of the resulting model.
One way to optimize a model for a given regression task is to consider multiple kinds of representations and build a weighted combination, and treat the weights as hyperparameters. 
This line of reasoning has been used to optimize the performance of a ML scheme to estimate the formation energy of small molecules~\cite{bart+17sa} and the chemical shieldings in molecular crystals.~\cite{paru+18ncomm}
Both applications  compound local descriptions with increasing cutoff spheres and decreasing weights, outperforming the best individual representation model.
The decaying weights assigned to representations with larger cutoffs reflect the multi-scale nature of the interactions that affect the values of chemical shieldings and of the molecular cohesive energy, which are often determined predominantly by the closest neighboring atoms and depend less markedly on atoms that are farther away.
To confirm this intuition \citet{will+18pccp} compare a similar mixture of representation with a radially scaled representations to model the formation energy of small molecules which corresponds to \cref{eq:trans-density} with
\begin{equation}
    \density{\alpha\br}{\CX_i} =  \sum_{j\in\CX_i,\alpha}  \; \gaus{\br -\brij}{2\sigma^2} f_c(r_{ij}) u(r_{ij}),
    \label{eq:radial-scaling}
\end{equation}
where $u(r_{ij})$ is a flexible radial scaling that reduces the weight of atoms in the far field.
The prediction accuracies of the mixture of representations and radial scaling model are shown to be very similar after independent optimisation of the model parameters with cross-validation.
This example showcases how incorporating physical insights about the target property into the representation helps in building more effective models.

Another scheme by which the atom-density framework can be generalized to reflect structure-property relations builds upon the similarities in the behavior of different chemical elements, which is reflected in the well-known trends observed along the periodic table.
Discarding such knowledge by considering each chemical species as completely different seems wasteful and even impractical when working on large subsets of the periodic table.~\cite{Artrith2017,Seko2015, Nyshadham2019}
Following this intuition, \citet{de+16pccp} formulated an `alchemical' kernel to supplement the SOAP power spectrum with the correlations between chemical species based on Pauling electronegativity. 
With the same mindset a distance across the periodic table has been proposed to learn properties across chemical composition space in the FCHL representation.~\cite{fabe+18jcp}

\begin{figure}[bhtp]
\begin{center}
    \includegraphics[scale=0.50]{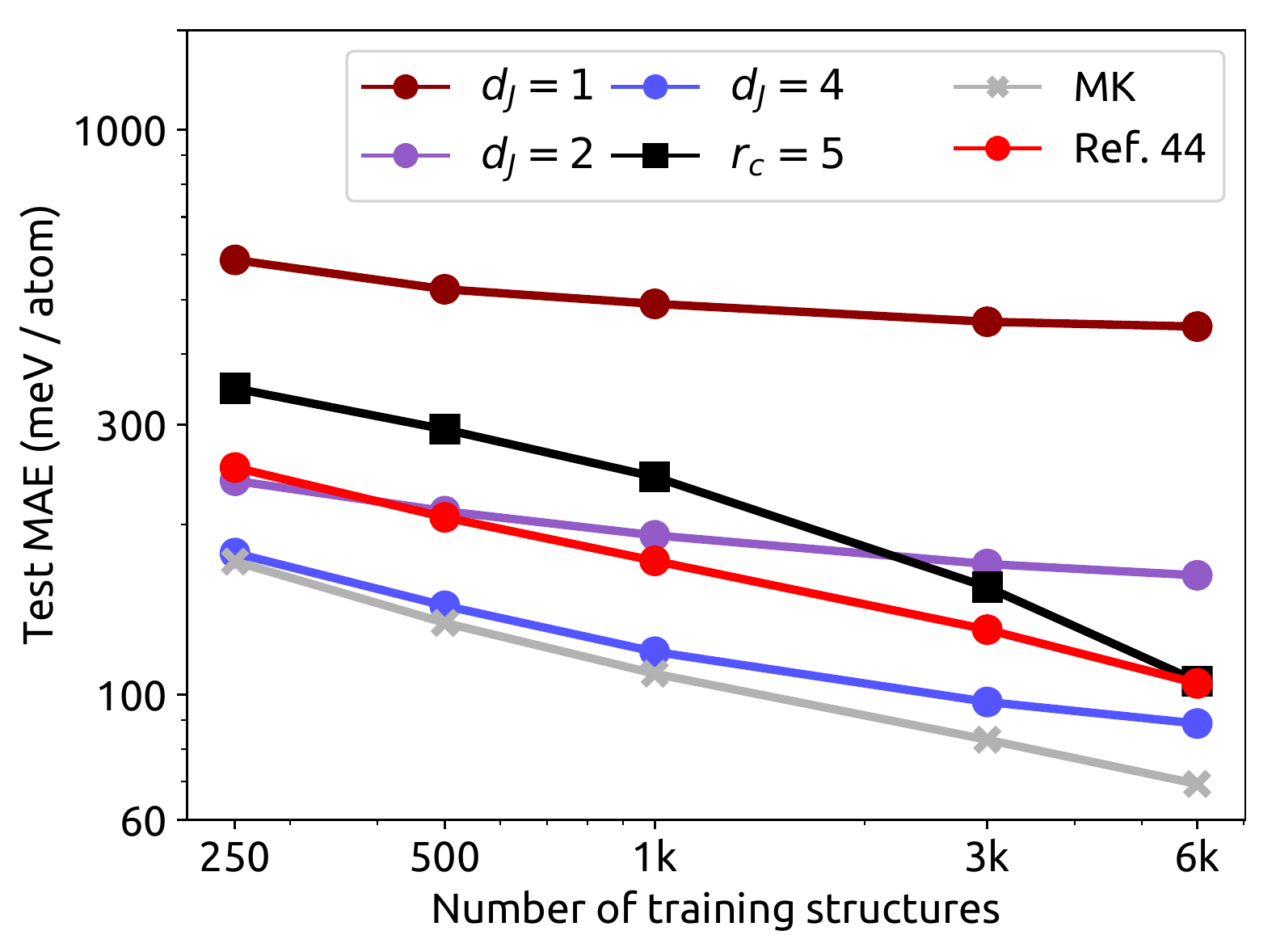}
\end{center}
\caption{
Learning curves for the formation energy of the elpasolite crystals database using GPR.~\cite{fabe+16prl} The standard power spectrum curve is shown in black, the best curve from Ref.~\citenum{fabe+18jcp} is shown in bright red and the optimized curves are shown in dark red ($d_J=1$), purple ($d_J=2$) and blue ($d_J=4$). For each of these models, the kernel was constructed with $r_{c}=5$\AA, $n_\text{max}=12$ radial basis functions and $l_\text{max}=9$ non-degenerate spherical harmonics. The compound model (shown in grey) combines three standard power spectrum representations and one chemically compressed representation ($d_{J}=4$, $r_{c}=5$) in the ratio $4:3:1:220$. Adapted from Ref.~\citenum{will+18pccp}.}
\label{fig:elpasolites-lc}
\end{figure}

Rather than using elemental properties to define \textit{a priori} the similarity between elements, the optimization of the representation of chemical space can be set as an additional objective of the ML algorithm. 
Then, the chemical features that characterize each element are learnt directly according to the dataset and target property at hand. 
Several applications of ANN to model the PES of molecules or solids use the stoichiometry as an input and the resulting features tend to match well with the structure of the periodic table.~\cite{Herr2019, Xie2018, Schutt2017} 
An alternating least square optimization procedure has been proposed to achieve similar results within the GPR framework.~\cite{will+18pccp}
It effectively corresponds to finding the best projection of the abstract elemental kets $\ket{\alpha}$ on an ``elemental feature'' basis $\ket{J}$, i.e. an embedding space, of dimension $d_J$ by optimizing the coefficients $u_{\alpha J} = \braket{J}{\alpha}$ within the modified power spectrum representation
\begin{equation}
    \sum_{\alpha\alpha'} u_{\alpha J}u_{\alpha' J'} \sum_m (-1)^m \density{\alpha nlm}{\CX_i}{} \density{\alpha'n'l -m}{\CX_i}{}.
    \label{eq:chem-opt}
\end{equation}
In \cref{fig:elpasolites-lc}, the learning curves of several chemically compressed models are compared with a baseline model, a compound model and the model taken from Ref.~\citenum{fabe+18jcp} for a chemically diverse benchmark dataset.
The compressed models tend to saturate because the low-dimensional ``elemental features'' are not sufficiently descriptive to account for the differences between the 39 elements in the dataset.
Nevertheless, in the limit of small training set size the compressed models (with $d_J=2,3,4$)  clearly outperform the baseline model.
The compound model (grey line) that combines both the baseline representation and the representation with $d_J=4$ avoids the saturation of the learning and retains the improved learning for small training set sizes.

\begin{figure*}[bhtp]
\centering
\includegraphics[width=1.0\linewidth]{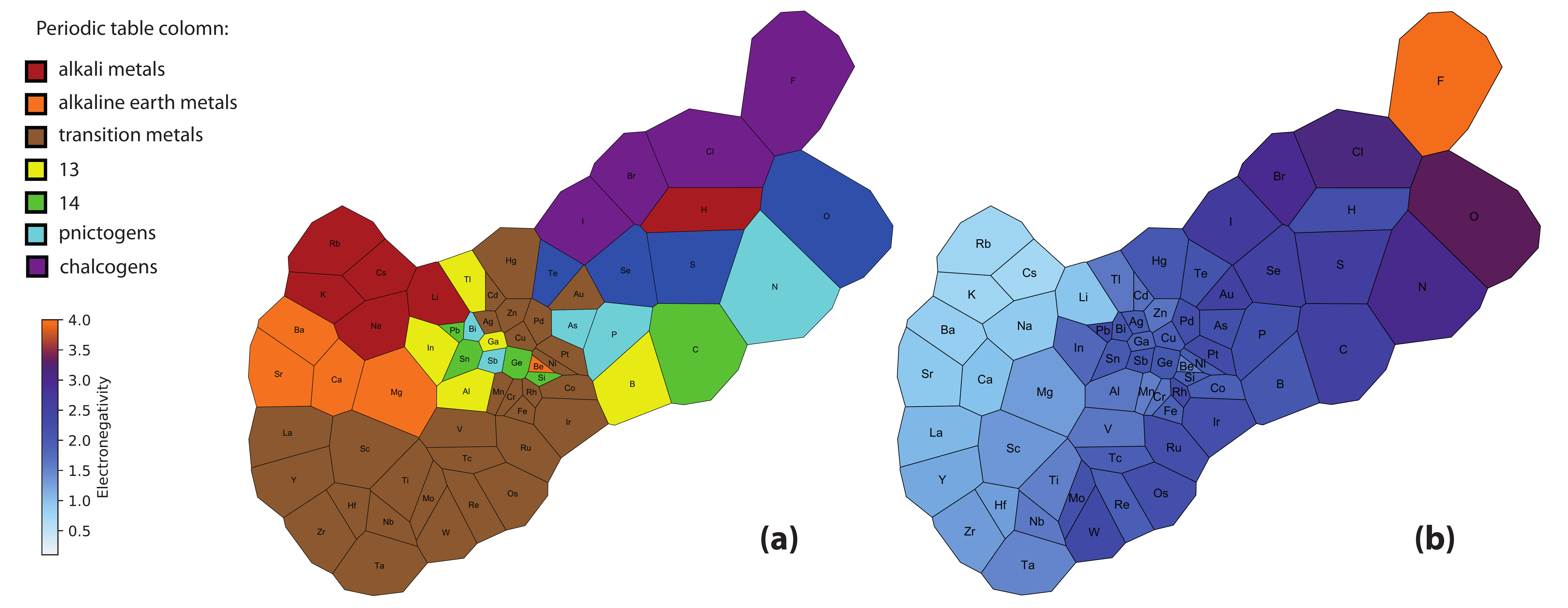}
\caption{Representation of the chemical space obtained by optimizing  a model of the lattice energy for a set of  perovskites~\cite{schm17c} whose composition is based on combinations of 64 elements. The map corresponds to the coefficients $u_{\alpha J}$ with $d_J=2$ (see \cref{eq:chem-opt}). Each element is represented with a Voronoi cell where the facets are at the midpoints between neighboring elements. The elements are color coded according to (a) their group in the periodic table and (b) Pauling electronegativity.}
\label{fig:pero-chem}
\end{figure*}

After optimization, the embedding space contains information on the similarity between elements with respect to the target property.
The ``elemental features'' obtained on a dataset of perovskites (62 different elements) and a model trained to predict  their formation energy is shown in \cref{fig:pero-chem} for $d_J=2$.
The resulting projection of the chemical elements evokes their positions in the periodic table which is highlighted by the coloring according to the periodic table group.
Moreover, the spatial arrangement of the two dimensional projection appears well correlated with the electronegativity.
Such data-driven techniques are powerful since they adapt to the system and target property but this also comes at an increased computational cost. 
Furthermore, the optimized chemical space might not be transferable across classes of systems, or for the learning of different properties. 

Besides data efficiency and the accuracy of predictions, numerical efficiency is also an essential criterion for a representation, since it affects the  length and time scales of problems that it can be used with. 
For example \citet{Caro2019} proposed an approximation to compute the SOAP power spectrum which results in a clear speedup with a marginal loss of accuracy. 
Similarly, the FCHL representation has been reformulated~\cite{Christensen2019} using much simpler functional forms increasing the numerical efficiency without impacting much the accuracy of the model.
In addition to improve the cost of computing the feature vectors associated with a given representation, computational effort can also be cut by reducing the number of features that need to be computed and used as input of the ML model.
ML schemes such as the CUR decomposition and the FPS scheme have been used  to  select a subset of the components of the power spectrum representation, and identify the most important parameters for Behler-Parrinello symmetry functions, obtaining simpler and more efficient models that were equivalent in performance to the full models.~\cite{imba+18jcp}

In closing, let us note that the Dirac notation that we have used to introduce the symmetrized atom-density framework also makes it possible to formulate many existing optimizations as the application of a linear operator that preserves the symmetries of the representation.~\cite{will+19jcp} 
For example a rotationally invariant operator that acts on the chemical part of a representation has matrix elements
\begin{equation}
    \matrixel{\alpha n l m}{\hat{U}}{\alpha' n' l' m'} = \delta_{nn'}\delta_{ll'}\delta_{mm'}\matrixel{\alpha}{\hat{U}}{\alpha'},
\end{equation}
when written in the same basis of radial functions and spherical harmonics used in~\cref{eq:sph-exp}. A low-rank expansion of such operator can be written as
\begin{equation}
    \hat{U} \approx \sum_{J\alpha} u_{J\alpha} \dyad{J}{\alpha},
\end{equation}
corresponding to a transformation of the chemical space into low-dimensional basis $\ket{J}$.
Applying such an operator to the power spectrum representation leads directly to \cref{eq:chem-opt}.
Similar operators can be formulated resulting in the geometrical scaling introduced by \citet{fabe+18jcp} or in the radial scaling of \cref{eq:radial-scaling}.
The atom density based framework is helpful to rationalize \textit{ad hoc} attempts to improve representations, generalize them and formulate rigorously more complex modifications that would, for instance, couple the geometry and the composition channels.

\section{Conclusion}

In the last decade ML techniques have demonstrated their utility in the context of atomistic simulations, both by automating the post-processing of large amounts of data, e.g. molecular dynamics trajectories, and by improving the efficiency and/or the accuracy of the prediction of atomic scale properties -- most notably through the construction of machine-learned interatomic potentials. 
Some consensus is starting to develop around the features of an effective ML model, e.g. the utility of incorporating symmetries and physical principles in the construction of representations of atomic neighborhoods, the importance of active-learning strategies and uncertainty quantification, and the the need to balance computational efficiency, data efficiency and transferability. %
Moreover, several software packages for building MLIPs and predicting atomic-scale properties have been made publicly available, and have been interfaced with efficient simulation codes, helping their dissemination beyond the groups which focus on methodological developments.~\cite{Bartok2015,Wang2018,Abbott2019,Schutt2019-chnet,Singraber2019,Gastegger2018,Khorshidi2016}
ML models have already been integrated in a few well established workflows of such as global structure optimization~\cite{Jorgensen2017} and crystal structure prediction.~\cite{Podryabinkin2019}
These applications, which need to generate and then screen thousands of configurations for a given system, and that do not need the models to be transferable across completely different compounds to yield a substantial acceleration, are the first candidates to benefit from ML. 
Methods that require sampling of fluctuations, such as molecular dynamics, are also very well suited to ML, as an effective interpolation scheme can cut down dramatically the cost needed to obtain accurate estimates of thermodynamic and time-dependent properties.~\cite{mora+16pnas,chen+19pnas}
One of the main challenges to make the use of ML more streamlined, and to obtain more transferable models, lies in efficiently generating a reference dataset, which requires the sampling of a large variety of configurations. 
Currently, many training sets for MLIPs have been built by sampling references from MD trajectories~\cite{Behler2016,Huan2017,li+15prl,Vandermause2019,Jinnouchi2019} which might seem wasteful since only a fraction of a trajectory is useful to the model.
Moreover, to achieve good performance on a wide range of thermo-mechanical properties, \citet{Bartok2018} had to design their training set by carefully introducing hand crafted configurations. 
Random structure search has been proposed to automate an efficient exploration of a wide range of configurations in solids~\cite{Deringer2018-boron,Bernstein2019} and active learning techniques to sample relevant configurations from already existing databases of isolated molecules~\cite{Smith2018} and molecular materials~\cite{paru+18ncomm} have shown some promise towards building more transferable models.

Optimizing the computational cost of obtaining ML predictions is another research direction that has been increasingly important, with several studies highlighting the tradeoffs between cost and accuracy.~\cite{Li2018,Nguyen2018,Kamath2018,Zuo2019} 
The extension of ML models beyond potentials, to predict more complex properties, from response tensors to quantum mechanical observables such as the electron charge density or the Hamiltonian is likely to become increasingly important.~\cite{gris+19acscs,Schutt2019,Schmidt2018,Zubatyuk2019,musi+18cs,wilk+19pnas,Wang2019,Christensen2019,broc+17nc,Ji2018}
As these efforts progress, making ML more useful and accessible, it will become clearer how the incorporation of ML  techniques in the atomistic modeling toolbox makes it possible to investigate scientific and technological problems that were inaccessible to electronic structure methods and empirical property models.

\section{Acknowledgments}
FM and MC were supported by the NCCR MARVEL, funded by the Swiss National Science Foundation.
The authors would also like to thank David Wilkins, Michael Willatt and Andrea Anelli for insightful comments on an early version of this manuscript.

\end{document}